\documentclass[prd,twocolumn,showpacs,preprintnumbers,amsmath,amssymb,superscriptaddress,floatfix,nofootinbib]{revtex4}

\usepackage{graphicx}
\usepackage{amsmath}
\usepackage{amsfonts}
\usepackage{amssymb}
\usepackage{color}
\usepackage{multirow}
\usepackage[colorlinks, citecolor=blue,anchorcolor=red,menucolor=red, linkcolor=red,filecolor=red,runcolor=red,urlcolor=blue,frenchlinks=red]{hyperref}

\begin{document}

\title{Strong decays of the higher isovector scalar mesons}

\author{Guan-Ying Wang}
\affiliation{Department of Physics, Zhengzhou University, Zhengzhou, Henan 450001, China}

\author{Shi-Chen Xue}
\affiliation{Department of Physics, Zhengzhou University, Zhengzhou, Henan 450001, China}

\author{Guan-Nan Li}
\affiliation{Department of Physics, Zhengzhou University, Zhengzhou, Henan 450001, China}

\author{En Wang\footnote{Corresponding Author: wangen@zzu.edu.cn}}
\affiliation{Department of Physics, Zhengzhou University, Zhengzhou, Henan 450001, China}

\author{De-Min Li\footnote{Corresponding Author: lidm@zzu.edu.cn}}
\affiliation{Department of Physics, Zhengzhou University, Zhengzhou, Henan 450001, China}

\vspace{0.5cm}

\begin{abstract}

Under the assignment of $a_0(1450)$ as the ground isovector scalar meson,
the strong decays of $a_0(1950)$ and $a_0(2020)$ are evaluated in the $^3P_0$ model.
Our calculations suggest that $a_0(1950)$ and $a_0(2020)$ can be regarded as the same resonance referring to $a_0(3^3P_0)$.
The masses and strong decays of $a_0(2^3P_0)$ and $a_0(4^3P_0)$
 are also predicted, which can be useful in the search for radially excited scalar mesons in the future.

\end{abstract}

\date{\today}

\maketitle

\section{Introduction}
\label{sec:intro}

In the framework of quantum chromodynamics (QCD), apart from the ordinary $q\bar{q}$ states,
other exotic states such as glueballs, hybrids, and tetraquarks are permitted
to exist in meson spectra. To identify these exotic states, one needs to distinguish them from the
background of ordinary $q\bar{q}$ states, which requires one to understand well the conventional $q\bar{q}$ meson
spectroscopy both theoretically and experimentally.

Experimentally, $a_0(980)$, $a_1(1260)$, $b_1(1235)$, and $a_2(1320)$, respectively, as the the lowest-lying $J^{PC}=0^{++}$, $1^{++}$, $1^{+-}$, and $2^{++}$ states, have been well established~\cite{PDG2016}. In contrast to $a_1(1260)$, $b_1(1235)$, and $a_2(1320)$, which can be well accommodated in the ordinary $q\bar{q}$ picture, $a_0(980)$, together with its multiplet partners $\sigma$, $\kappa$, and $f_0(980)$, does not fit well into the predictions of the quark model. For example, the observed mass
ordering of these lowest scalar states is $m_\sigma < m_\kappa < m_{a_0,f_0}$~\cite{PDG2016},
while in the conventional quark model, by a naive counting of the quark mass,
the mass ordering of the scalar $q\bar{q}$ nonet should be
 $m_\sigma \sim m_{a_0} < m_\kappa < m_{f_0}$. These scalar states below 1 GeV are generally believed not to be $q\bar{q}$ states~\cite{Close:2002zu,Maiani:2004uc,Amsler:2004ps,Pelaez:2003dy,Jaffe:2004ph,Eichmann:2015cra}.

At present, above the $a_0(980)$ mass, three higher isovector scalar states, $a_0(1450)$, $a_0(2020)$, and $a_0(1950)$, have been reported experimentally. $a_0(1450)$ was observed in $p\bar{p}$ annihilation experiments~\cite{Amsler:1994pz,Amsler:1995bf}, $D^\pm\rightarrow K^+ K^-\pi^\pm$~\cite{Rubin:2008aa}, and $D^0\rightarrow K^0_S K^\pm \pi^\mp$~\cite{Aaij:2015lsa}.
$a_0(2020)$ with two alternative solutions of similar masses and widths was found by the Crystal Barrel Collaboration in the partial wave analysis of the data on $\bar{p}p\to\pi^0\eta$
 and $\pi^0\eta'$~\cite{Anisovich:1999jv}, and $a_0(1950)$ was observed by the BABAR Collaboration in the processes  $\gamma\gamma\to K^0_sK^\pm \pi^\mp$ and  $\gamma\gamma\to K^+K^- \pi^0$~\cite{Lees:2015zzr}. The masses and widths of the three isovector scalar states are listed in Table~\ref{tab:scalar}. The lattice QCD calculations support that the lowest isovector scalar $q\bar{q}$ state corresponds to $a_0(1450)$ rather than $a_0(980)$~\cite{Mathur:2006bs,Burch:2006dg,Prelovsek:2004jp}. It is widely accepted that $a_0(1450)$ is the isovector member of the $1^3P_0$ $q\bar{q}$ nonet~\cite{PDG2016}. The natures of $a_0(2020)$ and $a_0(1950)$ are unclear. To be able to understand the nature of a newly observed state, it is natural and necessary to exhaust the possible $q\bar{q}$ description before restoring to more exotic assignments. Therefore, with the assignment of $a_0(1450)$ as the ground $q\bar{q}$ state, one naturally asks whether the higher isovector scalar states, $a_0(2020)$ and $a_0(1950)$, can be identified as the radial excitations of $a_0(1450)$.

\begin{table}[htpb]
\begin{center}\label{tab:scalar}
\caption{ \label{tab:fit}The masses and widths of the higher isovector scalar states (in MeV).}
\begin{tabular}{cccc}
\hline\hline
 State  & Mass  & Width & Ref.  \\
\hline
$a_0(1450)$ & $1474\pm19$ & $265\pm 13$ & \cite{Amsler:1994pz,Amsler:1995bf}  \\
  \multirow{2}{*} {$a_0(2020)$} & $2025\pm30$ & $330\pm75$ & \cite{Anisovich:1999jv} \\
& $1980^{+12}_{-80}$ & $225^{+120}_{-32}$ & \cite{Anisovich:1999jv} \\
$a_0(1950)$ & $1931\pm14\pm22$  & $271\pm22\pm29$ & \cite{Lees:2015zzr}  \\
\hline\hline
\end{tabular}
\end{center}
\end{table}

 Theoretical efforts on the quark model assignments for $a_0(2020)$ and $a_0(1950)$ have been carried out.
It is suggested that $a_0(1950)/a_0(2020)$ can be assigned as the $2^3P_0$ state based on the extended linear sigma model in Ref.~\cite{Parganlija:2016yxq}, where $a_0(2020)$ is considered earlier evidence for $a_0(1950)$.
In addition, $a_0(1950)/a_0(2020)$ is assigned as the $3^3P_0$ state based on the relativistic quark model in Ref.~\cite{Ebert:2009ub}, where the predicted $a_0(3^3P_0)$ mass is about 1993~MeV,
in agreement with both the $a_0(1950)$ and $a_0(2020)$ masses within errors. Obviously, further studies on the quark model assignments for $a_0(1950)$ and $a_0(2020)$ in other approaches are needed. Also, from Table~\ref{tab:scalar}, one can see that the resonance parameters of $a_0(2020)$ are close to those of $a_0(1950)$. The observed mass difference between $a_0(1950)$ and $a_0(2020)$ is less than 100~MeV; in such a small mass interval, it would be very difficult to accommodate two radial excitations of $a_0(1450)$ in practically all the quark models. We therefore conclude that if both $a_0(2020)$ and $a_0(1950)$ can be explained as $q\bar{q}$ states, they should correspond to the same resonance. In this work, we shall discuss the possible quark model assignments of $a_0(2020)$ and $a_0(1950)$ by investigating their strong decays in the $^3P_0$ model and check whether $a_0(2020)$ and $a_0(1950)$ can be identified as the same scalar meson.

The organization of this paper is as follows. In Sec.~\ref{sec:formalism}, we give a brief review of the $^3P_0$ model. In Sec.~\ref{sec:results}, the calculations and discussion are presented, and the summary and conclusion are given in Sec.~\ref{sec:summary}.

\section{THE $^3P_0$ MODEL }
\label{sec:formalism}

In this work, we employ the $^3P_0$ model to evaluate the Okubo-Zweig-Iizuka-allowed open flavor two-body  strong decays of the initial meson.  The $^3P_0$ model, also known as the quark-pair creation model, was originally introduced by Micu \cite{Micu:1968mk} and further developed by Le Yaouanc $et$ $al.$\cite{LeYaouanc:1972vsx,LeYaouanc:1973ldf,LeYaouanc:alo}. The $^3P_0$ model has been widely applied to study strong decays of hadrons with considerable success~\cite{Roberts:1992js,Blundell:1996as,Barnes:1996ff,Barnes:2002mu,
Close:2005se,Barnes:2005pb,Zhang:2006yj,Ding:2007pc,Li:2008mza,Li:2008we,Li:2008et,Li:2008xy,
Li:2009rka,Li:2009qu,Li:2010vx,Lu:2014zua,Pan:2016bac,Lu:2016bbk}. The main assumption of the $^3P_0$ model is that the strong decay occurs through a quark-antiquark pair with the vacuum quantum number. The new produced quark-antiquark pair, together with the $q\bar{q}$ within the initial meson, regroups into two outgoing mesons in all possible quark rearrangement ways.

Following the conventions in Ref.~\cite{Li:2008mza}, the transition operator $T$ of the decay  $A\rightarrow BC$ in the $^3P_0$ model is given by
\begin{eqnarray}
T=-3\gamma\sum_m\langle 1m1-m|00\rangle\int
d^3\boldsymbol{p}_3d^3\boldsymbol{p}_4\delta^3(\boldsymbol{p}_3+\boldsymbol{p}_4)\nonumber\\
{\cal{Y}}^m_1\left(\frac{\boldsymbol{p}_3-\boldsymbol{p}_4}{2}\right
)\chi^{34}_{1-m}\phi^{34}_0\omega^{34}_0b^\dagger_3(\boldsymbol{p}_3)d^\dagger_4(\boldsymbol{p}_4),
\end{eqnarray}
where the $\gamma$ is a dimensionless parameter denoting the probability of the quark-antiquark pair $q_3\bar{q}_4$ with quantum number $J^{PC}=0^{++}$. $\boldsymbol{p}_3$ and  $\boldsymbol{p}_4$ are the momenta of the created quark  $q_3$ and  antiquark $\bar{q}_4$, respectively. $\chi^{34}_{1,-m}$, $\phi^{34}_0$, and $\omega^{34}_0$ are the spin, flavor, and color wave functions of $q_3\bar{q}_4$, respectively. The solid harmonic polynomial  ${\cal{Y}}^m_1(\boldsymbol{p})\equiv|p|^1Y^m_1(\theta_p, \phi_p)$ reflects the momentum-space distribution of the $q_3\bar{q_4}$.

The partial wave amplitude ${\cal{M}}^{LS}(\boldsymbol{P})$ of the decay  $A\rightarrow BC$ can be given by~\cite{Jacob:1959at},
\begin{eqnarray}
{\cal{M}}^{LS}(\boldsymbol{P})&=&
\sum_{\renewcommand{\arraystretch}{.5}\begin{array}[t]{l}
\scriptstyle M_{J_B},M_{J_C},\\\scriptstyle M_S,M_L
\end{array}}\renewcommand{\arraystretch}{1}\!\!
\langle LM_LSM_S|J_AM_{J_A}\rangle \nonumber\\
&&\langle
J_BM_{J_B}J_CM_{J_C}|SM_S\rangle\nonumber\\
&&\times\int
d\Omega\,\mbox{}Y^\ast_{LM_L}{\cal{M}}^{M_{J_A}M_{J_B}M_{J_C}}
(\boldsymbol{P}), \label{pwave}
\end{eqnarray}
where ${\cal{M}}^{M_{J_A}M_{J_B}M_{J_C}}
(\boldsymbol{P})$ is the helicity amplitude and defined as,
\begin{eqnarray}
\langle
BC|T|A\rangle=\delta^3(\boldsymbol{P}_A-\boldsymbol{P}_B-\boldsymbol{P}_C){\cal{M}}^{M_{J_A}M_{J_B}M_{J_C}}(\boldsymbol{P}).
\end{eqnarray}
$|A\rangle$, $|B\rangle$, and $|C\rangle$ denote the mock meson states defined in Ref.~\cite{Hayne:1981zy}.

Due to different choices of the pair-production vertex, phase space convention, and employed meson space wave function, various $^3P_0$ models exist in the literature. In this work, we employ the simplest vertex as introduced originally by Micu, who assumes a spatially  constant pair-production strength $\gamma$\cite{Micu:1968mk}, relativistic phase space, and simple harmonic oscillator (SHO) wave functions. With the relativistic phase space, the decay width
$\Gamma(A\rightarrow BC)$ can be expressed in terms of the partial wave amplitude,
\begin{eqnarray}
\Gamma(A\rightarrow BC)= \frac{\pi
|\boldsymbol{P}|}{4M^2_A}\sum_{LS}|{\cal{M}}^{LS}(\boldsymbol{P})|^2, \label{width1}
\end{eqnarray}
where $|\boldsymbol{P}|=\frac{\sqrt{[M^2_A-(M_B+M_C)^2][M^2_A-(M_B-M_C)^2]}}{2M_A}$,
and $M_A$, $M_B$, and $M_C$ are the masses of the mesons $A$, $B$, and $C$, respectively. The explicit expressions for  ${\cal{M}}^{LS}(\boldsymbol{P})$ can be found in Refs.~\cite{Li:2008mza,Li:2008we,Li:2008et}.

Under the SHO approximation, the meson space wave function in the momentum space is
\begin{eqnarray}
\psi_{nLM_L}(\boldsymbol{p})=R_{nL}^{\text{SHO}}(p)Y_{LM_L}(\Omega_p),
\end{eqnarray}
where the radial wave function is given by
\begin{eqnarray}
R_{nL}^{\text{SHO}}(p)=&&\frac{(-1)^n(-i)^L}{\beta^{3/2}}\sqrt{\frac{2n!}{\Gamma(n+L+3/2)}}\nonumber\\
&&\times\left(\frac{p}{\beta}\right)^Le^{-(p^2/2{\beta}^2)}L_n^{L+(1/2)}\left(\frac{p^2}{\beta^2}\right).
\end{eqnarray}
Here $\beta$ is the SHO wave function scale parameter, and $L_n^{L+(1/2)}\left(\frac{p^2}{\beta^2}\right)$ is an associated Laguerre polynomial.

\section{CALCULATION AND RESULTS}
\label{sec:results}

In our calculations, the model parameters include the light nonstrange quark pair creation strength $\gamma$, the SHO wave function scale $\beta$, and the constituent quark masses. $\beta$ is set to be $\beta_A=\beta_B=\beta_C=0.4$ GeV, the typical values used to evaluate the light meson decays, as in Refs.~\cite{Barnes:1996ff,Barnes:2002mu,Li:2008mza,Li:2008we,Li:2008et,Li:2008xy,Li:2009rka,Pan:2016bac,Blundell:1995ev,Blundell:1995au,Ackleh:1996yt}, and the constituent quark masses are taken to be $m_u=m_d=330$~MeV and $m_s=550$~MeV, as in Refs.~\cite{Close:2005se,Li:2008mza,Li:2008we,Li:2008et,Li:2008xy,Li:2009rka,Pan:2016bac}.
 We take $\gamma=7.1$ by fitting to the total width of $a_0(1450)$ as the $1^{3}P_{0}$ state. The strange quark pair creation strength $\gamma_{s\bar{s}}$ can be related by $\gamma_{s\bar{s}}=\gamma\frac{m_u}{m_s}$~\cite{LeYaouanc:1977gm}.
The meson flavor wave functions follow the conventions of Refs.~\cite{Barnes:2002mu,Godfrey:1985xj} except for
$f_1(1285)=-0.28 n\bar{n}+0.96 s\bar{s}$, $f_1(1420)=-0.96 n\bar{n}-0.28 s\bar{s}$, as in Ref.~\cite{Li:2000dy}, $\eta(1295)=(n\bar{n}-s\bar{s})/\sqrt{2}$,
$\eta(1475)=(n\bar{n}+s\bar{s})/\sqrt{2}$, as in Ref.~\cite{Yu:2011ta}, where $n\bar{n}=(u\bar{u}+d\bar{d})/\sqrt{2}$. Masses of the final state mesons are taken from~\cite{PDG2016}.

The decay widths of $a_0(1450)$ as the $1^3P_0$ state are listed in Table~\ref{tab:fit}. The dominant decay modes of the $1^3P_0$ isovector state are $\pi\eta$, $\pi\eta'$, and $K\bar{K}$, consistent with observations of $a_0(1450)$~\cite{Amsler:1994pz,Amsler:1995bf,Uehara:2009cf}.

\begin{table}[htpb]
\begin{center}
\caption{ \label{tab:fit}Decay widths of $a_0(1450)$ as the $1^3P_{0}$ state (in MeV). The initial state mass is set to be 1474~MeV}
\begin{tabular}{ccc}
\hline\hline
 Channel                         & Mode            & $\Gamma_i(1^3P_0)$  \\
\hline
 $0^+\rightarrow 0^-0^-$         & $\pi\eta$         & 72.77    \\
                                  & $\pi\eta'$     & 144.47      \\
  $ $                             & $\pi\eta(1295)$  & 0.86          \\
  $ $                             & $K\bar{K}$             & 34.40               \\
  $0^+\rightarrow 0^-1^+$        &$\pi b_{1}(1235)$     &9.96          \\
                                  &$\pi f_{1}(1285)$    &3.98          \\
  Total width     & &$266.45$    \\
Experiment~\cite{PDG2016} & & $265\pm 13$ \\
\hline\hline
\end{tabular}
\end{center}
\end{table}

\begin{figure}[htpb]
\includegraphics[scale=0.7]{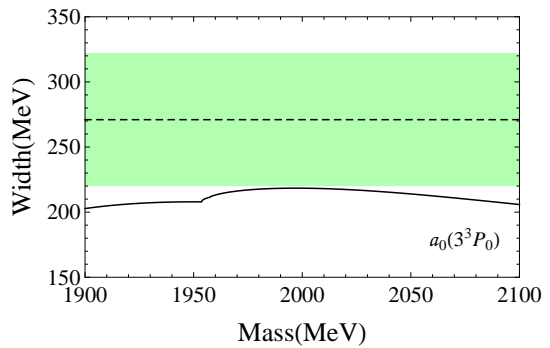}
\vspace{0.0cm}\caption{The dependence of the total width of $a_0(3^3P_0)$ on the initial state mass. The dashed line with a green band denotes the BABAR experimental data~\cite{Lees:2015zzr}.}\label{fig:1950twidth}
\end{figure}

\begin{figure}[htpb]
\includegraphics[scale=0.7]{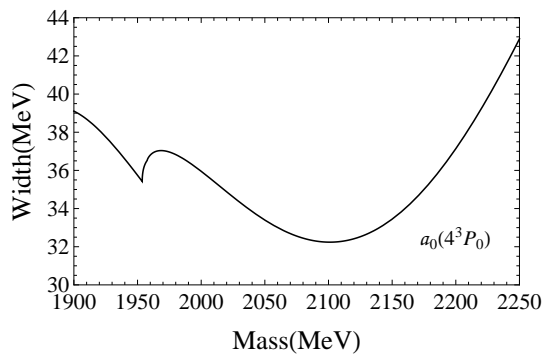}
\vspace{0.0cm}\caption{The dependence of the total width of $a_0(4^3P_0)$ on the initial state mass.}\label{fig:43p0twidth}
\end{figure}

The decay widths of $a_0(1950)$ as the $2^3P_0$ and $3^3P_0$ states are  shown in Table~\ref{tab:1950}. If $a_0(1950)$ is the $2^3P_0$ state, its total width is expected to be about 771 MeV, much larger than the observed $a_0(1950)$ width of $271\pm 22\pm 29$~MeV~\cite{Lees:2015zzr}. The possibility of $a_0(1950)$ being the $2^3P_0$ state can be ruled out. If $a_0(1950)$ is the $3^3P_0$ state, its total width is about 207~MeV, reasonably close to the measurement within errors. The dependence of the total width of $a_0(3^3P_0)$ on the initial state mass is shown in Fig.~\ref{fig:1950twidth}. Within the $a_0(1950)$ mass errors, the total width does not change too much. The assignment of $a_0(1950)$ as the $4^3P_0$ state can also be ruled out because the predicted width for $a_0(4^3P_0)$ with a mass of 1931 MeV is about 37.3 MeV (see also Fig.~\ref{fig:43p0twidth}), much smaller than the $a_0(1950)$ width. Therefore, the measured mass and width for $a_0(1950)$ are in favor of it being the $3^3P_0$ state.

\begin{table}
\begin{center}
\caption{ \label{tab:1950}Decay widths of $a_0(1950)$ as the $2^3P_0$ and $3^{3}P_0$ states (in MeV). The initial state mass is set to be 1931~MeV.}
\begin{tabular}{cccc}
\hline\hline
 Channel                      & Mode            & $\Gamma_i(2^3P_0)$   & $\Gamma_i(3^3P_0)$ \\
\hline
 $0^+\rightarrow 0^-0^-$      & $\pi\eta$         & 26.00  &5.13 \\
                              & $\pi\eta'$     &2.07 &2.52      \\
  $ $                         & $\pi(1300)\eta$   &37.75 &29.72           \\
  $ $                         & $\pi\eta(1475)$    & 20.81    &13.22            \\
  $ $                         & $\pi\eta(1295) $   &7.03 &1.97  \\
  $ $                         & $K\bar{K}$         &0.42 &0.74  \\

  $0^+\rightarrow 0^-1^+$     &$\pi b_{1}(1235)$     &339.48    &90.06      \\
                              &$\pi f_{1}(1285)$    &84.42       &9.18   \\
                              &$\pi f_{1}(1420)$    &4.38        &0.61  \\
                              &$KK_{1}(1270)$    &28.31          &9.97 \\
                              &$KK_{1}(1400)$    &2.38          &0.72\\
                              &$\eta a_{1}(1260)$    &24.12        &3.74  \\

  $0^+\rightarrow 1^-1^-$     &$\rho\omega$     &181.18       &36.07   \\
                              &$K^*\bar{K}^*$    &10.94         &2.05 \\
      $0^+\rightarrow 0^- 2^-$            &$\pi\eta_2(1645)$    &1.24  &1.58        \\

 Total width &    & $770.51$ &207.27    \\

 Experiment~\cite{Lees:2015zzr}&  & \multicolumn{2}{c}{$271\pm 22\pm 29$ }\\
\hline\hline
\end{tabular}
\end{center}
\end{table}

 As shown in Fig.~\ref{fig:1950twidth}, $a_0(3^3P_0)$ with a mass of 2025 MeV is predicted to have a width of about 217 MeV , about 38 MeV smaller than the lower limit of Crystal Barrel's solution I for the $a_0(2020)$ width of $330\pm 75$ MeV~\cite{Anisovich:1999jv}, and the predicted width for $a_0(3^3P_0)$ with a mass of 1980 MeV is about 218~MeV, in agreement with Crystal Barrel's solution II for the $a_0(2020)$ width of $225^{+120}_{-32}$ MeV~\cite{Anisovich:1999jv}. The possibility of $a_0(2020)$ being the $2^3P_0$ state can be ruled out because the expected width for $a_0(2^3P_0)$ with a mass of 1980 (2025) MeV is about 895 (995) MeV, much larger than the observed width of $a_0(2020)$, as shown in Table~\ref{tab:scalar}. The predicted width for $a_0(4^3P_0)$ with a mass of 1980 (2025) MeV is about 36.8 (34.6) MeV (see also Fig.~\ref{fig:43p0twidth}), much smaller than the $a_0(2020)$ width, which makes $a_0(2020)$ unlikely to be the $4^3P_0$ state. So, the measured mass and width for $a_0(2020)$ are consistent with an assignment of the $3^3P_0$ state.

The experimental evidence for both $a_0(1950)$ and $a_0(2020)$ turns out to be consistent with the presence of the same resonance corresponding to $a_0(3^3P_0)$. This naturally establishes 1.9 GeV as the approximate mass for the $n\bar{n}$ members of the $3P$ nonets, which could be useful to search for the $n\bar{n}$ members of the $3P$ nonets experimentally. The dominant decay modes of $a_0(3^3P_0)$ are $\pi(1300)\eta$, $\pi\eta(1475)$, $\pi b_1(1235)$, $KK_1(1270)$, and $\rho\omega$.

$a_1(1640)$ and $a_2(1700)$ as the $2P$ radial excitations have been established~\cite{Barnes:1996ff,Page:1998gz}, which also fixes the natural mass scale for the $n\bar{n}$ members of the $2P$ multiplets as about 1.7 GeV. One can expect to find $a_0(2^3P_0)$ near 1.7 GeV. At present, no candidate for the isovector scalar state around 1.7 GeV is reported experimentally. An $a_0$-like pole associated to a resonance with a mass of about 1760 MeV is found by investigating the meson-meson interaction in Refs.~\cite{GarciaRecio:2010ki,Geng:2008gx}. The $a_0(2^3P_0)$ mass in the extended linear sigma model is expected to be $1790\pm 35$ MeV~\cite{Parganlija:2016yxq}. Systematic studies on the meson spectra in the relativistic quark models show that the expected $a_0(2^3P_0)$ mass is about $1679\sim1780$ MeV~\cite{Ebert:2009ub,Godfrey:1985xj}.
Phenomenologically, it is suggested that the light mesons could be grouped
into the following Regge trajectories\cite{Anisovich:2000kxa},
\begin{equation}
M^2_n=M^2_0 + (n-1) \mu^2,
\label{regge}
\end{equation}
where $M_0$ is the lowest-lying meson mass, $n$ is the radial quantum number, and $\mu^2$ is the slope parameter of the corresponding trajectory.
In the presence of $a_0(1450)$ and $a_0(1950)/a_0(2020)$ being the $1^3P_0$ and $3^3P_0$ states, respectively, the $a_0(2^3P_0)$ mass
 can be determined to be about 1744 MeV based on Eq.~(\ref{regge}),\footnote{We take $M_{a_0(1450)}$=1474 MeV, $M_{a_0(1950)/a_0(2020)}=(1931+2025)/2$=1978 MeV, the average value of the $a_0(1950)$ mass reported by the BABAR Collaboration~\cite{Lees:2015zzr} and the favoured solution for the $a_0(2020)$ mass~\cite{Anisovich:1999jv}.}  consistent with the extended linear sigma model prediction~\cite{Parganlija:2016yxq} and the quark model predictions~\cite{Ebert:2009ub,Godfrey:1985xj}.

\begin{figure}[htpb]
\includegraphics[scale=0.7]{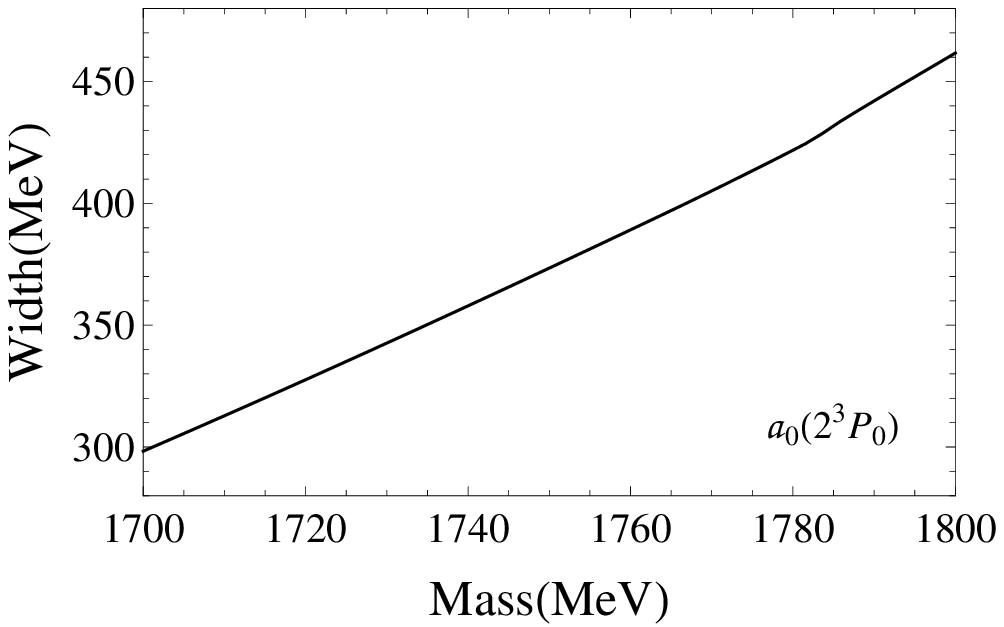}
\vspace{0.0cm}\caption{The dependence of the total width of $a_0(2^3P_0)$ on the initial state mass.}\label{fig:1760twidth}
\end{figure}

The strong decays of $a_0(2^3P_0)$ with a mass of 1744~MeV are presented in Table~\ref{tab:1760}. The total width of  $a_0(2^3P_0)$ is expected to be about 364~MeV. The dominant decay modes of $a_0(2^3P_0)$ include $\pi\eta(1475)$, $\pi\eta(1295)$, $\pi b_1(1235)$, $\pi f_1(1285)$, and $\rho\omega$.
The dependence of the total width of $a_0(2^3P_0)$ on the initial state mass is shown in Fig.~\ref{fig:1760twidth}. When the initial state mass varies from 1700 to 1800 MeV, the total width of the $a_0(2^3P_0)$ varies from about 298 to 460~MeV. With the initial state mass of 1700~MeV, our predicted width of 298 MeV is in agreement with the width of 293~MeV expected by Ref.~\cite{Barnes:1996ff} for $a_0(2^3P_0)$.

\begin{table}
\begin{center}
\caption{ \label{tab:1760}Decay widths of $a_0(2^3P_0)$ (in MeV). The initial state mass is set to be 1744~MeV.}
\begin{tabular}{ccc}
\hline\hline
 Channel                                & Mode            & $\Gamma_i(2^3P_0)$    \\
\hline
 $0^+\rightarrow 0^-0^-$          & $\pi\eta$                 & 7.91    \\
                                  & $\pi\eta'$             & 2.46     \\
  $ $                             & $\pi\eta(1475)$           & 19.25     \\
  $ $                             & $\pi\eta(1295) $          &20.86       \\
  $ $                             & $K\bar{K}$                      &1.07        \\

  $0^+\rightarrow 0^-1^+$         &$\pi b_{1}(1235)$          &213.08         \\
                                  &$\pi f_{1}(1285)$          &38.54           \\
                                  &$\pi f_{1}(1420)$          &1.02            \\

  $0^+\rightarrow 1^-1^-$         &$\rho\omega$          &59.96              \\

  \multicolumn{2}{c}{Total width}      & 364.12  \\
\hline\hline
\end{tabular}
\end{center}
\end{table}

\begin{table}[htpb]
\begin{center}
\caption{ \label{tab:43P0}Decay widths of $a_0(4^3P_0)$ (in MeV). The initial state mass is set to be 2187~MeV.}
\begin{tabular}{ccc}
\hline\hline
 Channel                 & Mode            & $\Gamma_i(4^3P_0)$    \\
\hline
 $0^+\rightarrow 0^-0^-$         & $\pi\eta$         & 0.00023    \\
                                  & $\pi\eta'$     &0.41       \\
  $ $                             & $\pi(1300)\eta$   &0.19       \\
  $ $                             & $\pi\eta(1475)$   & 0.77        \\
  $ $                             & $\pi\eta(1295) $  &4.63    \\
  $ $                             & $K\bar{K}$        &0.37     \\
  $ $                             & $KK(1460)$        &2.74     \\

  $0^+\rightarrow 0^-1^+$         &$\pi b_{1}(1235)$     &0.0032         \\
                                  &$\pi f_{1}(1285)$     &0.09        \\
                                  &$\pi f_{1}(1420)$     &0.04       \\
                                    &$KK_{1}(1270)$      &4.27      \\
                                     &$KK_{1}(1400)$     &2.50     \\
                                      &$\eta a_{1}(1260)$    &0.34         \\

  $0^+\rightarrow 1^-1^-$        &$\rho\omega$            &4.26        \\
                                   &$K^*\bar{K}^*$        &1.49      \\
          $0^+\rightarrow 0^- 2^-$                              &$\pi\eta_2(1645)$    &13.85           \\

 \multicolumn{2}{c}{Total width}     & $35.96$      \\
\hline\hline
\end{tabular}
\end{center}
\end{table}

$a_0(4^3P_0)$ is predicted to have a mass of about 2187 MeV based on Eq.~(\ref{regge}), consistent with 2250 MeV, the expected mass for $a_0(4^3P_0)$ in the quark model~\cite{Ebert:2009ub}.  The strong decays of $a_0(4^3P_0)$ with a mass of 2187 MeV are listed in Table ~\ref{tab:43P0}. The dependence of the total width of the $4^3P_0$ isovector state on the initial state mass is shown in Fig.~\ref{fig:43p0twidth}. A narrow width for $a_0(4^3P_0)$ is predicted. The $\pi\eta(1295)$, $KK_1(1270)$, $\rho\omega$, and $\pi\eta_2(1645)$ channels are the dominant decay modes for $a_0(4^3P_0)$. As we can see in Figs.~\ref{fig:1950twidth} and \ref{fig:43p0twidth}, the width derivatives are a discontinuity around 1950~MeV, which is because the decay channel $KK(1460)$ is open above this energy.

\section{SUMMARY AND CONCLUSION}
\label{sec:summary}

Observations of the state $a_0(1950)$ by the BABAR Collaboration have enlarged the family of the isovector scalar mesons. In this work, we discuss the possible quark model assignments of $a_0(1950)$ and $a_0(2020)$ by calculating their strong decays in the $^3P_0$ model. We suggest that $a_0(1950)$ and $a_0(2020)$ can be regarded as the same resonance referring to $a_0(3^3P_0)$. The confirmation of $a_0(1950)/a_0(2020)$ as the $3^3P_0$ state thereby establishes about 1.9 GeV as a natural mass scale for the $n\bar{n}$ members of the $3P$ nonets.

In the presence of $a_0(1450)$ and $a_0(1950)/a_0(2020)$ being the $1^3P_0$ and $3^3P_0$ states, respectively, in Regge phenomenology, the masses of $a_0(2^3P_0)$ and $a_0(4^3P_0)$ are predicted to be about 1744 MeV and 2187 MeV, respectively. The predicted masses for $a_0(2^3P_0)$ and $a_0(4^3P_0)$ are consistent with some other theoretical expectations. The total widths of $a_0(2^3P_0)$ and $a_0(4^3P_0)$ are expected to be about 364 MeV and 36 MeV, respectively. Our predictions could be useful to study the higher isovector scalar mesons experimentally.

\section*{Acknowledgements}
This work is partly supported by the National Natural Science Foundation of China under Grants No. 11505158 and No. 11605158, the China Postdoctoral Science Foundation under Grant No. 2015M582197, the Postdoctoral Research Sponsorship in Henan Province under Grant No. 2015023, and the Academic Improvement Project of
Zhengzhou University.

\end{document}